# The $^{139}$La($n,\gamma$) cross section: key for the onset of the $s$ process

R. Terlizzi[19], U. Abbondanno[20], G. Aerts[7], H. Álvarez[34], F. Alvarez-Velarde[30], S. Andriamonje[7], J. Andrzejewski[25], P. Assimakopoulos[16], L. Audouin[12], G. Badurek[1], P. Baumann[10], F. Bečvář[6], E. Berthoumieux[7], M. Calviani[18], F. Calviño[33], D. Cano-Ott[30], R. Capote[3,35], A. Carrillo de Albornoz[26], P. Cennini[36], V. Chepel[27], E. Chiaveri[36], N. Colonna[19], G. Cortes[32], A. Couture[40], J. Cox[40], M. Dahlfors[36], S. David[9], I. Dillmann[12], R. Dolfini[23], C. Domingo-Pardo[31], W. Dridi[7], I. Duran[34], C. Eleftheriadis[13], M. Embid-Segura[30], L. Ferrant[9], A. Ferrari[36], R. Ferreira-Marques[27], L. Fitzpatrick[36], H. Frais-Koelbl[3], K. Fujii[20], W. Furman[29], R. Gallino[22], I. Goncalves[27], E. Gonzalez-Romero[30], A. Goverdovski[28], F. Gramegna[18], E. Griesmayer[3], C. Guerrero[31], F. Gunsing[7], B. Haas[8], R. Haight[38], M. Heil[12], A. Herrera-Martinez[36], M. Igashira[24], S. Isaev[9], E. Jericha[1], Y. Kadi[36], F. Käppeler[12], D. Karamanis[16], D. Karadimos[16], M. Kerveno[10], V. Ketlerov[28,36], P. Koehler[39], V. Konovalov[29,36], E. Kossionides[15], M. Krtička[6], C. Lamboudis[13], H. Leeb[1], A. Lindote[27], I. Lopes[27], M. Lozano[35], S. Lukic[10], J. Marganiec[25], L. Marques[26], S. Marrone[a,19], C. Massimi[21], P. Mastinu[18], A. Mengoni[3,36], P.M. Milazzo[20], C. Moreau[20], M. Mosconi[12], F. Neves[27], H. Oberhummer[1], S. O'Brien[40], J. Pancin[7], C. Papachristodoulou[16], C. Papadopoulos[14], C. Paradela[34], N. Patronis[16], A. Pavlik[2], P. Pavlopoulos[11], L. Perrot[7], M. Pignatari[22], R. Plag[12], A. Plompen[5], A. Plukis[7], A. Poch[32], C. Pretel[32], J. Quesada[35], T. Rauscher[37], R. Reifarth[38], M. Rosetti[17], C. Rubbia[23], G. Rudolf[10], P. Rullhusen[5], J. Salgado[26], L. Sarchiapone[36], I. Savvidis[13], C. Stephan[9], G. Tagliente[19], J.L. Tain[31], L. Tassan-Got[9], L. Tavora[26], G. Vannini[21], P. Vaz[27], A. Ventura[17], D. Villamarin[30], M.C. Vincente[30], V. Vlachoudis[36], R. Vlastou[14], F. Voss[12], S. Walter[12], H. Wendler[36], M. Wiescher[40], K.Wisshak[12]

(n_TOF Collaboration)

[1]*Atominstitut der Österreichischen Universitäten,Technische Universität Wien, Austria,*
[2]*Institut für Isotopenforschung und Kernphysik, Universität Wien, Austria,*
[3]*International Atomic Energy Agency, NAPC-Nuclear Data Section, Vienna, Austria*
[4]*Fachhochschule Wiener Neustadt, Wiener Neustadt, Austria,*
[5]*CEC-JRC-IRMM, Geel, Belgium,*
[6]*Charles University, Prague, Czech Republic,*
[7]*CEA/Saclay - DSM, Gif-sur-Yvette, France,*
[8]*Centre National de la Recherche Scientifique/IN2P3 - CENBG, Bordeaux, France,*
[9]*Centre National de la Recherche Scientifique/IN2P3 - IPN, Orsay, France,*
[10]*Centre National de la Recherche Scientifique/IN2P3 - IReS, Strasbourg, France,*
[11]*Pôle Universitaire Léonard de Vinci, Paris La Défense, France,*
[12]*Forschungszentrum Karlsruhe GmbH (FZK), Institut für Kernphysik, Germany,*
[13]*Aristotle University of Thessaloniki, Greece,*
[14]*National Technical University of Athens, Greece,*
[15]*NCSR, Athens, Greece,*
[16]*University of Ioannina, Greece,*
[17]*ENEA, Bologna, Italy,*
[18]*Istituto Nazionale di Fisica Nucleare(INFN), Laboratori Nazionali di Legnaro, Italy,*
[19]*Dipartimento di Fisica and Istituto Nazionale di Fisica Nucleare, Bari, Italy,*
[20]*Istituto Nazionale di Fisica Nucleare, Trieste, Italy,*
[21]*Dipartimento di Fisica, Università di Bologna, and Sezione INFN di Bologna, Italy,*

---

[a] Corresponding author, address: via Orabona 4, 70125 Bari (Italy). Tel. +390805442511, e-mail stefano.marrone@ba.infn.it .

[22]*Dipartimento di Fisica, Università di Torino and Sezione INFN di Torino, Italy,*
[23]*Università degli Studi Pavia, Pavia, Italy,*
[24]*Tokyo Institute of Technology, Tokyo, Japan,*
[25]*University of Lodz, Lodz, Poland*
[26]*Instituto Tecnológico e Nuclear(ITN), Lisbon, Portugal,*
[27]*LIP - Coimbra & Departamento de Fisica da Universidade de Coimbra, Portugal,*
[28]*Institute of Physics and Power Engineering, Kaluga region, Obninsk, Russia,*
[29]*Joint Institute for Nuclear Research, Frank Laboratory of Neutron Physics, Dubna, Russia,*
[30]*Centro de Investigaciones Energeticas Medioambientales y Technologicas, Madrid, Spain,*
[31]*Istituto de Física Corpuscolar, CSIC-Universidad de Valencia, Spain,*
[32]*Universitat Politecnica de Catalunya, Barcelona, Spain,*
[33]*Universidad Politecnica de Madrid, Spain,*
[34]*Universidade de Santiago de Compostela, Spain,*
[35]*Universidad de Sevilla, Spain,*
[36]*CERN, Geneva, Switzerland,*
[37]*Department of Physics and Astronomy - University of Basel, Basel, Switzerland,*
[38]*Los Alamos National Laboratory, New Mexico, USA,*
[39]*Oak Ridge National Laboratory, Physics Division, Oak Ridge, USA,*
[40]*University of Notre Dame, Notre Dame, USA.*

Abstract

The nuclear resonance parameters and the capture cross section of the neutron magic isotope $^{139}$La has been measured relative to $^{197}$Au in the energy range of 0.6 eV-9 keV at the neutron time-of-flight (n_TOF) facility at CERN. The description of the experimental apparata is followed by the data analysis procedures concerning mainly the efficiency correction by means of the Pulse Height Weighting Function technique and the determination of different background components. We extracted the resonance parameters, the main nuclear quantities such as the resonance integral, $RI = 10.8 \pm 1.0$ barn, the average gamma widths for *s*- and *p*-waves, $\langle\Gamma_\gamma\rangle_{l=0} = 50.7 \pm 5.4$ meV and $\langle\Gamma_\gamma\rangle_{l=1} = 33.6 \pm 6.9$ meV, the nuclear level densities $\langle D\rangle_{l=0} = 252 \pm 22$ eV, $\langle D\rangle_{l=1} < 250$ eV, and the neutron strength functions $S_0 = (0.82 \pm 0.05)\times 10^{-4}$, $S_1 = (0.55 \pm 0.04)\times 10^{-4}$. These results represent a significant improvement over previous data. The deduced Maxwellian-averaged capture cross sections are important for the interpretation of the most recent spectroscopic observations in low metallicity stars.

# I. Introduction

Isotopes with closed neutron shells such as $^{139}$La ($N = 82$) are of special importance in nuclear physics. Their nuclear structure is characterized by low level densities and a small strength of the reaction channels, resulting in a low total neutron cross that is dominated by the elastic channel. The cross sections and the nuclear structure of these nuclei provide useful information for fundamental nuclear physics studies (microscopic calculations of many-body systems [1]) as well as for applications in nuclear astrophysics and nuclear technology. Concerning the latter aspect, $^{139}$La is an abundant fission fragment (5% fission yield [2]) and is widely used for neutron dosimetry in nuclear power plants because of the relatively long half-life ($t_{1/2}$ = 1.678 d) of $^{140}$La [3].

Lanthanum assumes a relevant role in nuclear astrophysics. It is copiously produced by the *main* component of the slow neutron capture process (*s* process) but receives also a sizable contribution from the rapid neutron capture process (*r* process). The dominant isotope $^{139}$La (99.91% in solar La) belongs to the second *s*-process peak shaped by the $N$=82 nuclei from Ba to Nd and is particularly suited for monitoring the *s*-process abundances from Ba up to Pb. Moreover, it is relatively easy to observe in stellar spectra, because transition probabilities and hyperfine structure constants of several lanthanum levels have been accurately measured [4]. Together with updated stellar atmosphere models and the use of high-resolution spectra, the lanthanum abundance can be reliably determined in stars of different metallicities. These analyses depend sensitively on the Maxwellian-average neutron capture cross section (MACS) at the typical temperatures of the astrophysical site of the *main s* process.

Unlike lanthanum, the close-by element europium is currently chosen as the best monitor of the *r* process. Europium is composed of two isotopes, $^{151}$Eu and $^{153}$Eu with solar isotopic abundances of 47.8% and 52.2%, respectively. The *s*-process contribution to solar europium is small (~6%), because both isotopes have very high neutron capture cross sections. The spectroscopic determination of the europium abundance is based on a selected number of lines with accurately measured hyperfine structure constants[5]. Accordingly, europium is a good marker of the *r*-process abundance distribution. In particular, correlated observations of lanthanum and europium abundances at various metallicities can provide an effective monitor of the chemical evolution in the Galaxy.

Time of flight (TOF) measurements of the $^{139}$La($n,\gamma$) cross section have been performed at Oak Ridge [6] and JAERI [7], and a transmission measurement was reported from the Columbia University synchrotron [8]. Recently, several activation measurements [9,10,11], have provided evidence for substantial systematic uncertainties in the previous TOF data. The serious astrophysical consequences resulting from discrepancies of up to 40% among the experimental data motivated a new capture cross section measurement at the innovative neutron time-of-flight facility n_TOF at CERN. The main features of the n_TOF installation such as the long flight-path, the extremely high instantaneous neutron flux, and the low background conditions together with an optimized detection set-up provided an ideal basis for accurate TOF measurements in a wide energy range, both for radioactive samples as well as for isotopes with very low capture cross sections, such as $^{139}$La.

In the following, the main characteristics of the facility and the experiment are described in Sec.II, the data analysis procedures in Sec.III, and the resulting resonance parameters in the resolved region from 0.6 eV up to 9 keV are reported in Sec.IV. The implications for fundamental nuclear physics and for nuclear astrophysics are eventually discussed in Sec. V.

## II. Experimental Apparatus

The main characteristics of the facilities, of the experimental apparatus, and of the data analysis procedures have been published in detail [12,13,14,15]. Therefore, we furnish here, together with a general description, only the specific features related to the lanthanum measurement.

**II.1 The n_TOF Facility**

At n_TOF, neutrons are generated via spallation by the 20 GeV/$c$ protons from the CERN PS accelerator complex impinging onto a massive target of natural lead. The measuring station is located at 187.5 m from the spallation target, inside the tunnel housing the evacuated neutron flight path. Because of the prolific neutron production provided by spallation reactions and of the very intense beam of $7\times10^{12}$ protons/pulse, the instantaneous neutron flux at n_TOF is more than two orders of magnitude higher than at other facilities. Backgrounds due to charged particles and $\gamma$-rays originating from the spallation target are efficiently reduced by several meters of concrete and iron shieldings, a sweeping magnet, and a combination of two collimators. The collimators serve for shaping of the neutron beam and for shielding of scattered neutrons from the target. The signal generated by the remaining fraction of ultrarelativistic particles (the so-called $\gamma$-flash) can be used for defining the start of the time-of-flight measurement. For the capture cross section measurements at n_TOF, an aperture 19 mm in diameter is used for the second collimator in front of the experimental area. This results in a neutron beam with an approximately Gaussian profile of 11.75 mm FWHM at the sample position. The main features of the n_TOF facility are summarized in Table I; for more technical details see Refs. [12][13][14].

**II.2 Detectors and Data Acquisition**

The prompt capture $\gamma$-rays were detected with two $C_6D_6$ liquid scintillation detectors with an active volume of ~1000 cm$^3$ [16]. The scintillator is contained in a thin-walled carbon fiber cell, which is directly coupled to an EMI 9823 QKA phototube without any further structural material around in order to minimize the sensitivity to sample scattered neutrons. The detectors are positioned 9.8 cm upstream of the sample with the front being about 3 cm from the beam axis as indicated by the schematic sketch of the setup in Fig. 1. The samples were mounted on a remotely controlled sample changer made from carbon fiber, which is directly integrated in the vacuum of the neutron beam line. Up to five samples can be mounted on the internal sample ladder for periodic background and reference measurements (see below).

The set-up for the capture measurements is complemented by the Silicon Flux Monitor (SiMon). The flux monitor consists of a thin $^6$Li deposit on a thin Mylar foil surrounded by a set of four silicon detectors outside the neutron beam [17] for recording the tritons and $\alpha$-particles from $^6$Li$(n,\alpha)^3$H reaction. This device is used for monitoring the neutron flux and to provide the normalization of the count rates measured with the different samples.

Due to the high instantaneous neutron flux, several events are generally recorded for a single neutron bunch. In order to avoid pile up and dead time problems, a data acquisition system based on high-frequency flash analog to digital converters (FADC) has been developed at n_TOF [18]. The FADC modules can be operated with sampling rates up to 1 GSample/s and are equipped with 8 MByte of buffer memory for each channel. The raw data are recorded signal by signal for detailed off-line analysis, which allows one to extract the required information on timing, charge, amplitude, and particle identification.

**II.3 Samples**

The characteristics of the samples are summarized in Table II together with the number of protons used for the measurement of each sample. In addition to the La under investigation, a gold reference sample as well as a natural C and a $^{208}$Pb sample were used in the present measurement. The La sample consisted of a natural metal foil (99.91% $^{139}$La and 0.09% $^{138}$La) enclosed in a low mass aluminum canning. The Au reference sample is included since the gold capture cross section is known with good accuracy, in particular in the region around the resonance at 4.9 eV and between 10 and 200 keV [19]. This sample is, therefore, used for neutron flux normalization [20]. The C sample and the $^{208}$Pb sample served for determining the background components related to sample scattered neutrons and to in-beam γ–rays.

## III. Data Analysis

The main steps in data analysis consist of the efficiency correction by means of the pulse height weighting technique (PHWT), followed by the determination and subtraction of the different background components and by the absolute normalization of the neutron flux. These aspects and minor additional corrections for dead time, self-shielding, and multiple scattering are discussed in this section.

**III.1 Pulse Height Weighting Technique**

Because of their low efficiency, the $C_6D_6$ detectors are detecting normally only a single γ-ray of the capture cascade. The probability of detecting a capture event depends, therefore, on the multiplicity of the cascade as well as on the energy of the emitted γ-rays, because of the intrinsic efficiency of the liquid scintillator. With the PHWT [21] the detector response $R(E_n, E_D)$, being $E_n$ the neutron energy and $E_D$ the energy deposited in the scintillator, is modified in such a way that the detection efficiency becomes independent of the cascade properties, but is completely determined by the neutron separation energy [15]. For obtaining the weighting function required for this correction, $WF(E_D)$, a set of response functions is calculated for each sample by detailed Monte Carlo simulations using the GEANT-3.21[22] and MCNP[23] software packages. In these simulations, the geometry and materials of the experimental set-up were carefully implemented and the points of origin of the photons were modeled according to the Gaussian neutron beam profile. The weighting functions are then defined by fitting the simulated response functions with a polynomial of degree four (see Equation 3 in the Ref. [15]).

The efficiency correction has to be corrected for coincidence events and for dead time. Coincident detection of two γ-rays has to be corrected because the PHWT is based on the assumption that only one γ-ray per capture cascade is recorded. In general, this condition, is ensured by the low overall efficiency of the liquid scintillator. In case of $^{139}$La, it is also supported by the low average cascade multiplicities and the low binding energy of 5.71 MeV. The probability for coincidences can be determined by the probability for coincidences between the two detectors, which was 2.1% for the adopted coincidence window of 20 ns. Since the corresponding correction for the Au reference sample was found to be 2.3% the true correction on the yield for La was less than ~ 1% with a relative uncertainty of 0.1%.

In principle, the dead time of the n_TOF data acquisition system is zero, but events can be reliably separated only if they are separated by more than 20 ns. This causes a virtual dead time that has estimated by means of the "Paralyzable Model" [24] approximation. In $^{139}$La measurement, this correction is negligible, except on top of the largest resonances (i.e. at 72 eV), where it reaches only ~ 0.5%.

With these corrections, the capture yield, $Y_{Raw}$, can be expressed by means of the PHWT [21]:

$$E_{casc} \int \Phi(E_n') Y_{Raw}(E_n') dE_n' = \sum_{E_D=0.2\ \text{MeV}}^{10\ \text{MeV}} R(E_n, E_D) WF(E_D), \qquad (1)$$

where $\Phi(E_n)$ is the total neutron fluence seen by the sample and $E_{casc}$ is the total capture energy converted to the laboratory frame. This raw yield has to be further corrected for the different background components as well as for other effects such as Doppler broadening, self-shielding, multiple scattering, and the neutron energy resolution. The raw capture yields of $^{nat}$La, carbon and lead samples are shown in Fig. 2.

**III.2 Background**

As reported in Ref. [25], the capture measurements at n_TOF are mainly affected by the "ambient" background, and by the sample-related contributions (in-beam γ-rays and scattered neutrons). The ambient background is mostly generated by the effect of the sample canning and by particles produced in the spallation target or in the collimators, which somehow reach the experimental area and produce signals in the capture set-up. This component, which was estimated by means of an empty aluminum container identical to that enclosing the lanthanum sample, is relatively low with respect to the lanthanum yield (Fig. 2).

A quantitative estimate of the background due to in-beam $\gamma$-rays was obtained for each sample by scaling the contribution measured with a $^{208}$Pb sample by means of detailed Monte Carlo simulations. These simulations were performed for $^{nat}$La, $^{197}$Au and $^{208}$Pb with the GEANT-3.21 package by using a detailed software model of the experimental apparatus. The respective $\gamma$-ray spectra were obtained by FLUKA simulations of the spallation and moderation process [26]. The $^{208}$Pb sample is particularly suited for this correction because it is very sensitive to $\gamma$-rays (high atomic number) and has a very low capture cross section (doubly magic). The effect of in-beam $\gamma$-rays in the $^{208}$Pb spectra are determined by subtracting the spectrum of the Al can. Both spectra were corrected for the weighting functions of the sample under consideration (La and Au). The resulting normalization factors $k_\gamma$ given in Table III.

The background from neutrons scattered by the sample and captured in or near the detectors was estimated by means of the carbon sample, because carbon is transparent to in-beam $\gamma$-rays and has a cross section that is dominated by the elastic channel. The scattering background is determined by subtracting the spectrum of the Al can (Fig. 2) and by rescaling the resulting yield with the normalization factor $k_n$, which is determined by the ratio between the scattering effect of the $^{nat}$La sample and of the carbon sample (Table III). This correction was found to be a few percent below 100 eV but negligible at higher neutron energies.

As illustrated in Fig. 2, the lanthanum yield is clearly above the total background up to 9 keV. Below 1 keV, the background is determined by the ambient component, whereas the contribution from in-beam γ-rays dominates in the upper part of the spectrum. According to the previous considerations, the background subtracted capture yield is:

$$Y_{BS}(E_n) = Y_{sample}(E_n) - Y_{Al-can}(E_n) - k_\gamma \left[ Y_{Pb}(E_n) - Y_{Al-can}(E_n) \right], \qquad (2)$$

where $Y_{sample}$ is the raw capture yield measured for the sample under study (La or Au), $Y_{Al-can}$ and $Y_{Pb}$ are the yields measured with the Al and $^{208}$Pb samples, respectively. All capture yields are consistently calculated using the same set of the weighting functions. The normalization for the total neutron fluence is obtained by the means of the SiMon neutron monitor. Other corrections concerning the neutron flux and the scattered neutrons are described below.

**III.3 Neutron Flux Determination**

Different detectors and techniques are employed at n_TOF to determine the total neutron flux. The most important and complete results are obtained with the calibrated fission chamber from PTB Braunschweig, with the SiMon, and with the analysis of standard resonances in the capture reactions of $^{197}$Au, Ag and $^{56}$Fe. The combination of these measurements yields the experimental neutron flux with an accuracy of better than 2% [12][13][14]. However, since the sample diameter is smaller than the neutron beam profile, only a fraction of this beam interacts with the capture samples. To estimate this fraction, the total flux is normalized according to the standard $^{197}$Au(n,γ) cross section, which is well known in the keV region and is considered standard for the resonance at 4.9 eV.

In the resolved resonance region from 1 to 100 eV, the flux fraction is calculated by fitting the main $^{197}$Au resonances with the *R*-matrix code SAMMY [27]. In the unresolved region, the normalization factor is evaluated by dividing, bin per bin, the gold cross section measured at n_TOF with the gold reference cross section [19]. The respective fractions of the beam seen by the samples, $C_{flux}$, are listed in Table IV for different neutron energies together with the results from FLUKA simulations. These values are in agreement with beam profile measurements performed with a Micromegas [20] detector. All experimental data and simulations confirm that the beam fraction $C_{flux}$ increases with neutron energy and must, therefore, be considered in the determination of the capture yield. Where available, the experimental values of $C_{flux}$ are used in the analysis (e.g. in the regions 1 – 100 eV and 10 – 100 keV). In the remaining energy regions, the results from the simulations were normalized to the experimental values.

**III.4 Other Corrections**

The determination of the resonance parameters requires that several minor effects have to be taken into account. These are the Doppler broadening of the resonance widths due to the thermal motion, the energy resolution of the neutron beam, the isotopic contamination of the sample, and the self-shielding and the multiple scattering effects in the sample.

All these corrections are included in $R$-matrix fits with SAMMY, which is used to extract the resonance parameters listed in Table V. The Doppler broadening is implemented according to the free-gas model with a temperature $T$=300 K and dominates over other sources of broadening below 1 keV neutron energy. The resolution function of the n_TOF neutron beam [28] becomes important above 1 keV neutron energy, where it describes the increasingly asymmetric resonance shape.

The very small isotopic impurity of $^{138}$La contributes to the low energy part of the spectrum as illustrated in Fig. 3. It is taken into account in the resonance fit by the corresponding resonance parameters from the JENDL-3.3 nuclear data library.

The self-shielding and multiple scattering corrections are an integral part of the standard SAMMY implementation. Considering the low total cross section of the lanthanum, a sizeable modification of the resonance parameters is produced only for the largest resonances, where the probability of elastic scattering is high. Generally, these effects modify the shape of the resonances by depressing the resonance peaks and overestimating the tails towards higher energy. These corrections affect the capture yield by a few percent with uncertainties of less than 1%.

## IV. Results

### IV.1 Resonance Parameters

The capture cross section of $^{139}$La is expressed in terms of *R*-matrix resonance parameters calculated in the Reich-Moore approximation with the code SAMMY [27]. The fit of the resonances, see Figs. 3 and 4, is performed in different ways in order to check the reliability of the extracted parameters. In general, the three resonance parameters $E_R$, $\Gamma_\gamma$ and $\Gamma_n$ are left free to vary while, according to the discussion in the previous section, the normalization factor is kept fixed. The main steps of the fitting procedure consisted in the extraction of the resonance parameters from the background-subtracted capture yields. The spin assignment of each level is carefully checked by comparing the fits obtained with different assignments indicated in the ENDF/B-VI.8, JEFF-3.1 and JENDL-3.3 libraries [29]. For many levels especially when $\Gamma_n \gg \Gamma_\gamma$, the best fit is found by adopting the transmission value as the neutron width, whereas in the few cases with $\Gamma_\gamma \gg \Gamma_n$, the $\gamma$-width is fixed to the average value (Table V). When the $\Gamma_n$ are left free to vary, our neutron widths are roughly consistent with the results of the transmission measurement [8]. The remaining differences might be attributed to the fact that several *p*-wave levels (in this work but especially in Hacken *et al.* [8]) are missing and that the neutron width does not always dominate the total width of the resonance but fluctuates following a Porter-Thomas distribution. In some cases the capture channel may therefore provide a sizable contribution to the total cross section while in other cases this contribution is negligible. This feature can also inferred from the trend of the *s*-wave and *p*-wave neutron strength function shown in Figs. 5 and 6.

In the thin sample approximation, the background-subtracted capture yields and the neutron capture cross section ($\sigma_\gamma$) are related by:

$$Y_{\text{Capture}}(E_n) = \left(1 - \exp^{-N_{\text{Atoms}}\sigma_{tot}(E_n)}\right) \frac{\sigma_\gamma(E_n)}{\sigma_{tot}(E_n)} \simeq \frac{N_{\text{Atoms}} C_{\text{flux}}(E_n)}{C_{\text{MS}}(E_n)} \sigma_\gamma(E_n), \tag{3}$$

where $\sigma_{tot}$ is the total neutron cross section, $N_{\text{Atoms}}$ is the number of atoms per barn, and $C_{\text{MS}}$ is the self-shielding and multiple scattering corrections. In this relation the Doppler broadening and the resolution function of the neutron beam are not considered. Both effects are included in the SAMMY fit together with the contribution of the potential scattering, which is calculated theoretically using a radius equal to $R' = 5.1$ fm [30]. This full analysis was checked by determining the resonance parameters from fits of the experimental data prior to background subtraction. On average, the parameters obtained in this way are within uncertainties consistent with those derived before.

In our full analysis three new resonances have been identified, and improved parameters were obtained for the previously known resonances. The final results extracted from the background-subtracted spectra are listed in Table V. As shown in Sec. IV.2, the fitting procedure allowed us to evaluate the statistical and systematic uncertainties of the lanthanum capture cross section in a coherent way. The results of some fits in the resolved resonance energy region up to 9 keV are illustrated in Figs. 3 and 4. According to the level assignment indicated in the Ref. [30], some neutron capture resonances of the $^{138}$La impurity (0.09%) can be seen in Fig. 3 at 3, 20, 67, and 89 eV neutron energy.

Up to 9 keV, all levels reported by Hacken et al. [8] and Musgrove et al. [6] could be identified. The three new resonances are *p*-waves at neutron energies of 6.036, 6.766, and 7.783 keV. In general, the orbital angular momentum assignments by Hacken et al. on the basis of the Bayes statistical analysis are confirmed, while most of the additional resonances detected by Musgrove et al. are *p*-waves. Up to 2 keV, the resonances energies ($E_R$) are in good agreement with previous results, but differences appear for resonances between 6 and 8 keV. It seems that the accuracy of the resonance energies has been overestimated by Hacken *et al.*, who took only the flight path contribution into account but did not include the neutron beam resolution (see Ref. [31]).

The comparison of the capture strengths ($g\Gamma_n \Gamma_\gamma / \Gamma_{tot}$) derived from the n_TOF data with previous results indicates that our capture strength is on average about 10% lower than reported by Nakajima *et al.*[7] and a few percent lower than those of Musgrove *et al.*[6]. A possible explanation for these systematic differences can be due to a more reliable evaluation of the PHWT, an accurate treatment of the corrections for self shielding, multiple scattering and the effect of neutron energy resolution, and the use of the well tested *R*-matrix code SAMMY. Finally, systematic uncertainties are significantly reduced by the improved n_TOF set-up, which exhibits much lower neutron sensitivity by the use of low-mass carbon fiber cells for the liquid scintillator detectors. In addition, the use of FADCs provides an efficient way for $n/\gamma$ - discrimination [32], and the extremely low repetition rate of one neutron burst every 2.4 s avoids pulse overlap and provides a low neutron induced background. It is important to note that these features provide possible explanations of the systematic differences between our results and previous data, but that there is no clear experimental evidence for the real causes. In particular, there seems to be no correlation between $\Gamma_n$ and $\Gamma_\gamma$ for the whole set of resonances.

In order to compute the MACS and the Resonance Integral, the present data have to be complemented to cover the range up to 1 MeV in neutron energy. The comparison with the evaluated cross sections from the JENDL, ENDF and JEF nuclear data libraries [29] shows that the n_TOF results agree best with JENDL in terms of orbital angular momentum assignments and of general resonance properties. The better agreement could be due to the fact that the JENDL evaluation takes all previous measurements [6,7,8] into account. Nevertheless, the capture cross sections measured at n_TOF are lower than the JENDL data, by about 10% in the first part of the neutron energy spectrum (<1 keV) and by 3% on average between 5 and 9 keV.

As illustrated in the next section, the main nuclear input for determining the lanthanum abundance in stellar model calculations are the MACS for a range of stellar temperatures between $kT = 5$ and 100 keV, which are obtained by folding the capture cross section with a Maxwell-Boltzmann type neutron spectrum in a wide energy range (100 eV-500 keV). Since the n_TOF results cover only part of this spectrum, only the partial contribution to P-MACS can be accurately determined as indicated in column two of Table VI. In view of the remaining differences between the n_TOF data and JENDL-3.3, the accurate experimental result of the MACS measured at $kT=25$ keV with the activation method by O'Brien et al. [10] at FZK has been considered. The MACS, calculated at $kT=25$ keV using the n_TOF data and the JENDL cross section, is 5% higher with respect to the FZK MACS. In order to reproduce the measured values of n_TOF and FZK, the JENDL cross section is renormalized with respect to the n_TOF data between 9 and 15 keV (by a factor of 0.97) and at higher energies with respect to the FZK MACS (by a factor of 0.95). The MACS uncertainties are calculated propagating the errors of the capture cross section used in the folding procedure. According to the previous discussion, the systematical uncertainty on the evaluated nuclear data is estimated around 10%.

A confirmation that this procedure is reasonable and the experimental data are accurate, comes from recent activation measurements at FZK [10,11]. Unlike the case of a prompt capture $\gamma$-ray measurement, the activation technique is not affected by background from scattered neutrons and is, therefore, particularly suited for the measurement low capture cross sections. The Maxwellian averaged capture cross section (MACS) evaluated by O'Brien et al. [10] at a thermal energy of $kT = 30$ keV (31.6 ± 0.8 mbarn) is 37% lower than the MACS obtained by Musgrove et al. [6] (50 ± 5 mbarn). The most recent result of the MACS, performed with the activation method by Winckler et al. [11] at $kT = 5$ keV (113.7 ± 4.9 mbarn) is within the quoted uncertainty consistent with the n_TOF value of 106.9 ± 5.3 mbarn.

Using the same approach, the Resonance Integral, defined as:

$$RI = \int_{0.5 \text{ eV}}^{1 \text{ MeV}} \sigma_\gamma(E_n)/E_n \, dE_n,  \qquad (4)$$

is calculated to be 10.8 ± 1.0 barn, which is 8.5% lower than the 11.8 ± 0.8 barn given in Ref.[30] but only 4.5% lower than a previous measurement (11.2 ± 0.5 barn) [33]. This result is consistent with the observation that the capture strengths have been found smaller with respect to the previous measurements [6,7] especially at low energy. The small difference between the experimental $RI$ [33] and the n_TOF value can be ascribed to the uncertainty of the thermal capture cross section, which is not measured in this work.

**IV.2 Uncertainty Analysis**

In the resolved resonance region, the total uncertainties related to the resonance parameters are derived from the $R$-matrix fit performed with the SAMMY code. The uncertainties (1 σ) of the parameters are listed in Table V. The contribution of each component to the total uncertainty is discussed in the following, and a summary of the error balance is reported in Table VII.

*Statistical:* Due to the low capture cross section of lanthanum, the statistical error represents the most important contribution to the final uncertainty. The statistical errors from the PHWT analysis are not simply given by the number of events in a given neutron energy bin, but have to be calculated by propagating the errors in the definition of the yields in Eq. (1) [34]. The total background is increasing with neutron energy and becomes comparable to the cross section at 9 keV, thus adding also a significant contribution. For the chosen energy binning, the sum of the contributions from the gold sample (see next paragraphs), from the La sample, and from the background, ranges from 4 to 8%.

*Weighting Function:* The overall uncertainty due to the PHWT has been determined by the comparison of the capture yields extracted with different sets of weighting functions. In fact, the weighting functions obtained with different codes or using different thresholds or a different polynomial order (third or fourth) in the least squares method, lead to differences of up to 3% in the extracted yield. These discrepancies are reduced to less than 1.5%, if the cross section is measured relative to a reference sample (i.e. Au), provided that the weighting functions are consistently calculated for both samples. In this case, the systematic effects (geometrical details, tracking of the photons, etc…) affect both samples similarly and cancel out to a large part. The remaining 1.5% uncertainty is a realistic value for the present measurement, because the uncertainty of this method was determined in Ref. [15] to be less than 2% from a complete analysis of the standard resonances in Au, Ag, and Fe.

*Neutron flux:* This systematic uncertainty is related to the shape of the absolute neutron flux that has been determined by the combination of different measurements [12] [13] [14] with an uncertainty of less than 2%. An additional contribution to this uncertainty is associated with the determination of the fraction of the beam seen by the sample, because the sample diameter was smaller than the beam size. As described before, the gold standard cross section was used to estimate the fraction $C_{\text{flux}}$ with a total uncertainty (statistical and systematic) of less than 4% (see Table IV). The combined fitting procedure together with the detailed simulations leads to an improved accuracy for these fractions of 2%. It is therefore reasonable to adopt a value of 2% for the systematic uncertainty of the flux normalization as indicated in Table VII.

*Minor Effects:* The number of atoms in each sample is very well determined by the mass measurement listed in Table II. Also the $^{138}$La impurity is well known by the natural composition of the sample. Therefore, the associated error is much less than 0.1% and was neglected. Uncertainties due to Doppler broadening, beam resolution, and self-shielding and multiple scattering corrections are estimated according to the total, elastic and capture cross sections calculated by SAMMY. These corrections and other minor effects contribute at most a few percent to the capture yield with average uncertainties of about 0.5%.

The total uncertainties of the resonance parameters are estimated on average to be of the order of 6.5% (see Tables V and VII). The PHWT and the flux determination constitute the main systematic contributions to those errors as reported in Table VII, while the uncertainties of the minor corrections are practically negligible except in the largest resonances. As a consequence of the low capture cross section of $^{139}$La, the largest uncertainty of this measurement is, therefore, due to the statistical contribution.

## V. Implications

### V.1 Nuclear Structure

The accurate resonance analysis presented for the neutron magic isotope $^{139}$La is of interest for several reasons. The level densities and the relative strength functions of the compound nucleus $^{140}$La can be derived for the different assignments of the orbital angular momentum. Since $^{139}$La has spin 7/2, the compound nucleus has two possible total angular momentum $J = 3$ and 4 for the thirty-nine $s$-wave resonances, four values $J = 2$, 3, 4, and 5 for the forty $p$-wave levels. The orbital angular momentum is deduced from the resonance fits, which take the previously known neutron widths into account. For the $s$-wave and $p$-wave set of levels the average spacing $\langle D \rangle$ calculated from the maximum likelihood fit of a Wigner distribution, the average gamma widths $\langle \Gamma_\gamma \rangle$, and the neutron strength function $S$ obtained by fitting the cumulative sum of the reduced neutron widths are summarized in Table VIII. The reduced neutron widths for $s$-waves are:

$$\Gamma_n^0 = \Gamma_n \left( \frac{1.0 \text{ eV}}{E_R} \right)^{1/2}, \tag{5}$$

and for the *p*-waves:

$$\Gamma_n^1 = \frac{1+(ka)^2}{(ka)^2} \Gamma_n^0, \tag{6}$$

where *k* is the neutron wave-number in the laboratory frame and *a* is the nuclear hard-sphere radius calculated according to the formula $a = 0.8 + 1.23A^{1/3}$ fm [29]. The cumulative number of levels and the cumulative sum of the reduced neutron widths for the *s*- and *p*-wave levels are shown in Figs. 5 and 6. All fits are performed for the entire neutron energy range (0.6 eV - 9 keV) except for the cumulative number of *p*-wave resonances. In this case, the energy range was restricted to energies below 2.5 keV because the sequence of the *p*-wave levels seems complete only up to this energy. Figs. 5 and 6 provide also important information concerning the nuclear structure of the compound nucleus. While the sequence of *s*-wave resonances seems to be complete up to 9 keV, the reduced neutron widths oscillate more than predicted by the Porter-Thomas (P-T) distribution, especially between 3 and 4 keV. This observation was already reported by Hacken *et al.* [8] for $^{139}$La, $^{140}$Ce, and $^{141}$Pr as a hint of undetected intermediate structures in these neutron magic nuclei. On the contrary, the good linearity of the cumulative sum of the neutron widths in Fig. 6 suggests that the strongest *p*-wave resonances are almost all detected and that the missing levels have very low capture strength.

In order to check the completeness of the level sequence, as well as the reliability of the orbital angular momentum assignment, statistical $\Delta_3$ tests [35] are performed for the *s*- and *p*-wave ensembles, and the number of missing levels is estimated via the P-T distribution of the reduced neutron widths $g\Gamma_n^l$ for $l = 0$ and 1. In particular, the theoretical $\Delta_3$ value is initially calculated according to the Dyson-Mehta ($\Delta_{3,\text{DM}}$) theory; see Eq. 81 in Ref. [35]. In order to have a further confirmation of this result and to assess more certain conclusions, we have simulated the behavior of the $\Delta_3$ statistic according to the Gaussian

Orthogonal Ensemble (GOE) matrices theory. Two sets of consecutive eigenvalues are extracted from two independent samples of 80,000 GOE matrices of dimension 300×300 randomly generated. From thirty-nine consecutive eigenvalues of each matrix the corresponding $\Delta_3$ value is calculated. Finally after having summed the two samples, the average value ($\Delta_{3,GOE}$) and the relative confidence region of the $\Delta_3$ distribution has been determined. According to those results, the statistical significance, i.e. the probability that $\Delta_3 \leq \Delta_{3,Exp}$, is 54% while it results smaller if a number of missing resonances is assumed. It has to be noticed also that $\Delta_{3,DM}$ and $\Delta_{3,GOE}$ are calculated for a level spacing ratio $D_{J=3}/D_{J=4} = 1.22$, although these results are quite insensitive with respect to this ratio. The experimental and theoretical $\Delta_3$ values are compared in Table VIII for the total number of $s$-waves. Although the number of levels of this ensemble is relatively poor (39), those results ($\Delta_{3,Exp}$, $\Delta_{3,DM}$, $\Delta_{3,GOE}$ and statistical significance) are compatible with the assumption that the set of $s$-wave levels is complete. For the $p$-wave resonances the standard deviation of theoretical $\Delta_3$ is much larger than the experimental value due to the four possible $J$-assignments. This feature induces large fluctuations in the cumulative number of $p$-wave levels and makes the relative statistical analysis ineffective.

Integration of the P-T distribution for a single level population provides the number of total levels, which can be calculated for $s$- and $p$-waves according to:

$$N(x) = N_l \left[1 - \mathrm{erf}\left(\sqrt{x/2}\right)\right], \qquad (7)$$

where $l = 0$ or $1$, $x = g\Gamma_n^l / \langle g\Gamma_n^l \rangle$, and $N_l$ is the total number of expected levels. The maximum likelihood fits of both distributions according to the Eq. (7) are illustrated in Fig. 7. The number of estimated $s$-wave levels is $N_0 = 39 \pm 3$, in perfect agreement with the experimental result, whereas $N_1 = 54 \pm 5$ indicates that we are missing 25% of the $p$-levels. However, the estimate for the $p$-wave resonances is only indicative since we were not able to determine $J$. In fact, the analysis of the $p$-wave ensemble was performed according to the P-T single channel distribution, but including all possible $J$-values. The correct approach [36] [37] would have been to treat the $J = 3$ and 4 states by means of a two channel P-T distribution ($p_{1/2}$ and $p_{3/2}$ neutron interactions), while the $J = 2$ and 5 levels can be described by a single channel P-T distribution ($p_{3/2}$). Nevertheless, the conclusions of this analysis support the previous observations derived from Figs. 5 and 6 and from the $\Delta_3$ tests.

**V.2 Nuclear Astrophysics**

Lanthanum plays an important role for the determination of the *s*-process abundances of the heavy elements. Together with the other isotopes of the second *s* peak around the magic neutron number $N = 82$, lanthanum acts as bottleneck between the abundant light *n*-capture elements Sr, Y, and Zr belonging to the first *s* peak at magic number $N = 50$ and the heavy elements from Sm up to Pb and Bi. Lanthanum is also a good marker in the interpretation of stellar spectra because it is practically monoisotopic and its absorption spectra contain several lines with accurately measured transition probabilities and hyperfine structure constants [4]. The remaining problem, partly solved by this work, has been the lack of reliable MACS values for the low temperature phase of *s*-process nucleosynthesis around $kT = 8$ keV.

Once the *s*-process abundance is accurately determined with respect to solar lanthanum, the *r*-process contribution is completely fixed by the residual, $N_r = N_\odot - N_s$. This decomposition is crucial for the understanding of chemically unevolved stars in the Galactic halo, which exhibit heavy element patterns that are essentially of *r* process origin produced by short-lived massive progenitor stars [38] [39]. Exceptions are the extrinsic Asymptotic Giant Branch (AGB) stars [40], which include the majority of C-rich stars [41] [42], the *s*-process rich stars [43] [44] [45], the Pb-rich stars [46], and other peculiar stars, like the *r*-process rich stars [47]. All these stars have quite different abundance distributions compared to the unevolved halo stars due to mass transfer in a binary system or due to the pollution by a nearby supernova. Therefore these stars must be excluded from the sample of unevolved halo stars. Note, however, that the literature data shown in Figs. 8 and 9 contain a few chemically peculiar stars, which have been added to the selected sample stars due to their unusual high [La/Eu]$^b$ ratio of >0.2 dex, the typical range predicted for the *s*-enhanced extrinsic AGB stars. A typical [La/Eu]$_s$ ~ 0.9 dex is indeed predicted in by the *s* process, and the observed for an extrinsic AGB depends on the degree of pollution.

Previously, barium was the preferred indicator for the heavy *s*-process elements in stellar spectra, because the capture cross sections of the related isotopes are all well known [48]. Barium has a relative large solar abundance among the heavy elements (4.43 relative to Si=10$^6$ in the solar system) [49] and the most part of it (~80%) is synthesized by the *s* process. Apart from the rare nuclei $^{130,\ 132}$Ba (which represent only 0.2% of solar Ba), barium is composed of five isotopes, from $^{134}$Ba to $^{138}$Ba. The two *s*-only isotopes $^{134,\ 136}$Ba contribute 10% to solar barium, while the neutron magic $^{138}$Ba, mostly of *s*-process origin, contributes 71.7%. Notice that the odd isotope $^{137}$Ba (11.2% of solar Ba) is also mainly of *s*-process origin [50]. From the spectroscopic point of view, however, barium presents several disadvantages. Barium exhibits only few spectral lines, one often saturated (4554 Å) and the others difficult to detect. In addition, the hyperfine splitting by the various isotopes contributes noticeably to a spread of strong lines. These features lead to relatively large uncertainties in the spectroscopy of barium and a corresponding scatter in the abundance determinations. Eventually, these uncertainties in the spectroscopic observations put the use of barium for monitoring the *s*-process contributions in stars into question. In contrast, lanthanum is not affected by such ambiguities, and is, therefore, considered as the more reliable *s*-process indicator.

Initially, the decomposition of the *s*-process abundance and the corresponding *r*-process residuals was based on the classical *s*-process approach [51], where three components (*main*, *weak* and *strong*) were invoked for describing the solar *s*-process distribution [52]. This phenomenological model has meanwhile led to inconsistencies in describing the *s* - abundances of the Nd isotopes and of the branching in the reaction path at $^{141}$Ce, $^{142}$Pr, $^{147}$Nd, $^{148}$Pm [44], and $^{151}$Sm [53], and was, therefore, essentially abandoned. The *main s*-component in the solar system is associated with nucleosynthesis processes in low mass stars (≤8 M$_\odot$) while evolving along the AGB and suffering recurrent thermal instabilities (Thermal Pulses, TP) in the He-shell (for a review see Busso, Gallino and Wasserburg [54]). After each TP, freshly synthesized $^{12}$C and *s*-process elements from between the He- and H-burning zone (the He-intershell) is mixed to the surface by third dredge-up episodes. Along the AGB phase, the star progressively loses its entire envelope by very efficient stellar winds, in this way returning *s*-enriched material to the interstellar medium.

---

$^b$ The spectroscopic notation indicates that: $[A/B] = \log_{10}(A/B) - \log_{10}(A/B)_\odot$ where A and B are elemental abundances. The symbol $\odot$ indicates that the abundance is referred to the solar value.

The major neutron source in AGB stars is provided by the $^{13}$C($\alpha$, *n*)$^{16}$O reaction, which operates at the top of the He intershell in the so-called $^{13}$C pocket. This phase is characterized by low neutron densities (~$10^7$ cm$^{-3}$) and low temperatures ($kT \approx 8$ keV) and operates between TPs under radiative conditions for periods of about $10^4$ yr [55] [56]. A second source of neutrons is provided by the $^{22}$Ne($\alpha$, *n*)$^{25}$Mg reaction, which is marginally activated at the high temperatures ($kT \approx 20$-$25$ keV) during TPs for a much shorter time interval of 5 to10 yr. Though high peak neutron densities up to $10^{11}$ cm$^{-3}$ are reached, the time integrated neutron flux is limited to about 5% of the total neutron budget. During a TP, the whole He-intershell region becomes convective. According to this scenario, the solar *main* component is the result of all previous generations of AGB stars in the Galaxy. As a matter of fact, the build up of the *s* elements is not a unique process but depends on the metallicity and initial stellar mass as well as on the strength of the $^{13}$C pocket and on the mass loss rate.

With the present MACS of $^{139}$La, the TP-AGB model of the *main* component yields an *s*-process contribution of 74 ± 3 % of the solar lanthanum abundance, very close to the values obtained by O'Brien *et al.* (76.9%) [10] and by Winckler *et al.* [11] (70.0%), consistently higher than reported by Arlandini *et al.* [50] (64.2%) on the basis of the previous MACS (see also Fig. 9). Our MACS, at $kT = 30$ keV, is identical to that of O'Brien *et al.*, but 6% lower at 5 keV compared to the result of Winckler *et al*. The small differences in the inferred *s* abundance of $^{139}$La are, in fact, due the MACS at low temperatures, because La is predominantly produced in the $^{13}$C pocket. While the minor neutron exposure by the $^{22}$Ne neutron source plays an important role for branching isotopes like $^{151}$Sm [57], which has a high MACS, it has almost no impact on the final abundance of neutron magic isotopes like $^{139}$La.

The improved understanding of the *s*-process production of lanthanum is important for Galactic chemical evolution. The *s* process is not of primary origin, because the necessary Fe seed does not result directly from hydrogen and helium burning in the same star. Consequently, the *s*-process abundance distribution depends on the stellar metallicity, on the initial mass, and on the strength of the carbon pocket [54]. Moreover, low mass AGB stars have a very long lifetime of several Gyr. Therefore, the *main* component can not contribute to the chemical enrichment of the early universe. In fact, the heavy *s* elements appear only at metallicities [Fe/H]>–2 [58,59,60]. Besides the *main* component, also the *strong s* component is the outcome of the *s*-process nucleosynthesis operating in AGB stars, as discussed in the references. In fact, recent studies [54,61,62] have identified in the low metallicity ([Fe/H]< –1.5) and low mass AGB stars the astrophysical origin for the synthesis of the *strong s* component. This production mechanism accounts for almost the 50% of solar $^{208}$Pb and the 20% of solar $^{209}$Bi. Only the *weak s* process, which occurs in the shorter-lived massive stars (>8 $M_\odot$), contributes the *s* nuclei up to $A \approx 90$ already at earlier times. This component is synthesized partly in the helium burning cores of massive stars and partly in the subsequent convective carbon burning shell.

Unlike the *s* process, the *r* process is believed to be of primary origin, in particular for the heavy elements beyond Ba. Possible astrophysical sites are type II supernovae and neutron stars mergers, which are principally suited to produce the extremely high neutron densities required to describe the *r*-process abundance distribution, but there are remaining open questions for both scenarios (for a review see Truran *et al.* [38]). Since it is occurring in massive stars (>8 $M_\odot$), which have short lifetimes of less than 0.1 Gyr, the *r* process dominates the production of heavy elements in the early Universe. This is confirmed by the observation of metal-poor halo stars, which exhibit clear *r*-process patterns. The analysis of TP-AGB stars has shown that the *s*-process contributions to europium is very low (~ 6%) and not affected by the branching of the *s* path in this mass region [59]. From the spectroscopic point of view, the europium abundance in stars is generally low (0.095 relative to Si=$10^6$ in the solar system) [49] but the Eu lines are well resolved and the respective hyperfine constants are accurately measured [5]. Therefore, the europium abundances can be reliably established, and even isotopic ratios have been determined by hyperfine splitting analyses of the spectral lines [63]. Due to these features, europium represents a good marker of the *r*-process abundance distribution.

In principle, the increase of the metal abundances is correlated with the age of the Galaxy, thus providing a sort of cosmo-chronometer. Metal-poor stars are particularly important in this respect, because they represent the oldest observable stellar population. According to the previous discussions, the *s* process can be characterized by lanthanum, the *r* process by europium, and the metallicity can be used as a "chronometer". Hence; the combination of these quantities permits to follow the chemical evolution of the Galaxy. In Figs. 8 and 9 the spectroscopic ratios [La/Fe], [Eu/Fe], and [La/Eu] are shown as a function of metallicity [Fe/H] for 227 stars in a metallicity range between 0 and –3. Most of these data are from Burris *et al.* [64] and Simmerer *et al.* [65], complemented by the *r*-process enhanced stars measured by McWilliam *et al.* [66], Johnson and Bolte [67], Honda *et al.* [68], and Barklem *et al.* [69]. The [element/Fe] ratio at low metallicity exhibits large variations of ± 1 dex (Fig. 8) because the iron production at low metallicity can not be considered an effective chronometer [70] [71] or because of inhomogeneities of the interstellar medium [39,72,73,74]. The large dispersion of the heavy elements in halo stars is also attributed to incomplete mixing of the interstellar medium in the first epochs of the Galaxy.

The results obtained in this work is most relevant with respect to the spectroscopic ratio [La/Eu] plotted in Fig. 9. Since the *main s*-process starts to contribute to the heavy element abundances for metallicities [Fe/H] ≥–1.5, the earlier lanthanum abundance should correspond to the much smaller *r*-process value, whereas the *r*-process element europium is produced at solar proportions. At low metallicities, the [La/Eu] ratio should, therefore, correspond to the *r*-process value of –0.56 ± 0.07 as indicated in Fig. 9 by the dashed-dotted line. The expected increase of the [La/Eu] at metallicities > –1.5 due to the later *s* contributions is clearly confirmed by the observational data.

The rise of the heavy element abundance from a pure *r*-process level due to the progressive enrichment by *s*-process elements emerged a few years ago thanks to observational [66,75,76] and theoretical [57,71] studies. While these arguments were first using the observed barium abundances[64,77,78,79], they are now relying more on lanthanum instead [65]. These studies have to be further pursued to clarify a number of open questions resulting from uncertainties in the observational data, in the nuclear aspects, and in the stellar models. For example, it is interesting to note that Johnson and Bolte [64] and Barklem *et al.* [65] do not find any rise of the [La/Eu] ratio in the entire range –3 < [Fe/H] < –1.5 over the *r*-process value (Fig. 4 in Ref. [67] and Fig. 20 in Ref. [69]).

In conclusion, the TOF measurement of the $^{139}$La($n,\gamma$) cross section has led improved MACS calculations at the low thermal energies, which are crucial for the determination of the *s*-process abundance of lanthanum. Nevertheless, the uncertainties in the astronomical observations both for the stars' abundances and for the solar abundances still remain large. In fact as illustrated in Fig. 8 and 9, the error on the spectroscopic ratios for each star is on average 0.2 dex. Moreover, in several cases the results of different observational groups are in disagreement between each other both for the metallicity values and for the heavy elements abundance estimates (see Table IX). For clarity, in Fig. 8 and 9 the error bars of the abundances of the single stars are not drawn whereas single stars observed by different authors are connected with a dashed line. Table IX lists also the stars whose abundances show striking variations ($\geq 0.2$ dex) between different observers. In addition, the shaded area in Fig. 9 represents the total error induced mainly by the solar abundance uncertainties (~15% corresponding to ±0.072 dex) estimated according to the data published in the Ref. [49]. It has to be remarked that a different compilation [80] of solar abundances dedicates a special treatment to the rare earth elements and estimates a lower uncertainty (~3% corresponding to ±0.021 dex). It is evident that this kind of uncertainties represents the major problem in the accurate determination of the *r*-process spectroscopic ratios. This situation should greatly benefit from the ongoing coordinated efforts in high resolution spectroscopy.

## VI. Conclusions

This paper reports the $^{139}$La($n,\gamma$) cross section measured at n_TOF. The capture cross section is given in terms of resonance parameters in a large energy range from 0.6 eV up to 9 keV. These results show sizeable differences with respect to the previous experimental data and allow one to extract the related nuclear quantities with improved accuracy. The results are important for the nuclear structure aspects of neutron magic isotopes, for nuclear technology, and particularly, the nucleosynthesis history of the Galaxy. The new capture cross sections have led to substantially improved Maxwellian averaged cross sections at the comparably low thermal energy of $kT = 8$ keV characteristic of the dominant *s*-process neutron source. This improvement contributed to the analysis of the La and Eu abundances over a wide range of stellar metallicity, which opens new vistas on the chemical evolution of the Galaxy.

## Acknowledgements

SM wishes to acknowledge J. J. Cowan for useful discussions concerning the elemental spectroscopic ratios. This work has been partly supported by the EC within the framework of the n_TOF-ND-ADS project (nuclear data for accelerator systems, contract FIKW-CT-2000-00107) and by the funding agencies of the participating institutes.

# Tables and Figures

**TABLE I. Main characteristics of the n_TOF facility.**

| Parameter | Comment |
| --- | --- |
| Proton beam | 20 GeV/$c$ momentum, 7 ns (rms) pulse width |
| | repetition rate 0.4 Hz (average) |
| | intensity $7\times10^{12}$ protons/pulse |
| Neutron beam | 300 neutrons/proton |
| | energy range from 0.6 eV to 250 MeV (set by the performance of the data acquisition system) |
| | one magnet, two collimators, and heavy Fe and concrete shielding for background reduction |
| | neutron filters for background definition |
| | $\sim 10^5$ neutrons/pulse/energy decade at 185 m distance from the spallation target |
| | neutron energy resolution $\Delta E/E = 10^{-3}$ at 30 keV |

**TABLE II. Sample characteristics and the relative number of protons used in the lanthanum measurement.**

| Sample | Diameter (mm) | Mass (g) | Number of protons [in units of $10^{17}$] |
| --- | --- | --- | --- |
| $^{nat}$La + Al can | 20 | 1.943 | 1.353 |
| Al can (empty) | 30 | 0.357 | 0.646 |
| $^{nat}$C | 20 | 3.933 | 0.423 |
| $^{208}$Pb | 20 | 4.941 | 0.551 |
| $^{197}$Au | 20 | 2.873 | 0.757 |
| *Total* | | | 3.730 |

**TABLE III.** Normalization factors for correction of backgrounds due to in-beam $\gamma$-rays ($k_\gamma$) and to neutrons scattered by the sample ($k_n$).

| Factor | $^{nat}$La | $^{197}$Au | $^{nat}$C | $^{208}$Pb |
|---|---|---|---|---|
| $k_\gamma$ | 0.335 | 0.665 | - | 1 |
| $k_n$ | 0.100 | - | 1 | - |

**TABLE IV.** The fraction of the neutron flux seen by the sample extracted from the gold data and from FLUKA simulations (adopted values).

| Energy range | 1-10 eV | 10-100 eV | 0.1-1 keV | 1-10 keV | 10-100 keV |
|---|---|---|---|---|---|
| $C_{flux}$ (experimental) | 0.480 ± 0.010 | 0.482 ± 0.010 | - | - | 0.504 ± 0.021 |
| $C_{flux}$ (simulated) | 0.480 | 0.483 | 0.488 | 0.493 | 0.500 |

**TABLE V.** Resonance parameters ($E_R$, $\Gamma_\gamma$ and $g\Gamma_n$), radiative capture strengths ($g\Gamma_n\Gamma_\gamma/\Gamma_{tot}$), and orbital angular momentum assignment ($l$) for the $n + {}^{139}$La system. Fixed values in the SAMMY fits are marked by an asterisk. The $g\Gamma_n$ values are adopted from Hacken *et al.* [8], while average $\gamma$-widths $\langle\Gamma_\gamma\rangle_l$ are used when $\Gamma_n \gg \Gamma_\gamma$.

| $E_R$ (eV) | $\Gamma_\gamma$ (meV) | $g\Gamma_n$ (meV) | $g\Gamma_n\Gamma_\gamma/\Gamma_{tot}$ (meV) | $l$ |
|---|---|---|---|---|
| $0.758 \pm 10^{-3}$ | $40.11 \pm 1.94$ | $5.6\cdot 10^{-5} \pm 5\cdot 10^{-6}$ | $5.6\cdot 10^{-5}$ | 1 |
| $72.30 \pm 0.05$ | $75.64 \pm 2.21$ | $11.76 \pm 0.53$ | 8.68 | 0 |
| $249.1 \pm 0.1$ | $33.6^*$ | $0.167 \pm 9\cdot 10^{-3}$ | 0.17 | 1 |
| $339.8 \pm 0.5$ | $33.6^*$ | $0.162 \pm 9\cdot 10^{-3}$ | 0.16 | 1 |
| $617.8 \pm 1.2$ | $36.35 \pm 1.58$ | $13.35 \pm 0.75$ | 8.08 | 0 |
| $703.8 \pm 1.9$ | $15.29 \pm 0.93$ | $6.20 \pm 0.19$ | 4.49 | 1 |
| $876.4 \pm 2.6$ | $25.45 \pm 1.24$ | $12.79 \pm 0.64$ | 6.75 | 0 |
| $906.2 \pm 3.1$ | $18.57 \pm 0.68$ | $4.49 \pm 0.19$ | 3.65 | 1 |
| $963.6 \pm 4.1$ | $29.48 \pm 1.53$ | $8.64 \pm 0.45$ | 5.17 | 0 |
| $1181 \pm 4$ | $91.62 \pm 2.73$ | $923^*$ | 38.41 | 0 |
| $1209 \pm 5$ | $19.73 \pm 1.55$ | $8.10 \pm 0.98$ | 4.68 | 1 |
| $1257 \pm 5$ | $26.13 \pm 3.50$ | $15.82 \pm 0.33$ | 6.64 | 1 |
| $1429 \pm 6$ | $33.6^*$ | $0.071 \pm 6\cdot 10^{-3}$ | 0.07 | 1 |
| $1433 \pm 6$ | $33.6^*$ | $2.15 \pm 0.12$ | 1.93 | 1 |
| $1639 \pm 7$ | $27.31 \pm 2.01$ | $43.9 \pm 6.5$ | 9.39 | 0 |
| $1652 \pm 7$ | $14.84 \pm 2.93$ | $32.3 \pm 5.9$ | 5.41 | 0 |
| 1828 ± 10 | 7.99 ± 1.05 | 5.25 ± 0.13 | 3.24 | 1 |
| 1918 ± 10 | 28.23 ± 2.71 | 4.63 ± 0.24 | 3.36 | 1 |
| 1973 ± 10 | 18.52 ± 2.75 | 2.21 ± 0.30 | 1.60 | 1 |
| 2119 ± 12 | 61.40 ± 4.70 | 1702* | 26.44 | 0 |
| 2155 ± 12 | 39.75 ± 2.75 | 480* | 21.36 | 0 |
| 2178 ± 12 | 48.39 ± 2.97 | 5.80 ± 0.35 | 4.55 | 1 |
| 2383 ± 14 | 51.10 ± 2.85 | 29.8 ± 2.1 | 12.77 | 0 |
| 2472 ± 14 | 62.53 ± 4.16 | 507* | 32.89 | 0 |
| 2667 ± 15 | 38.98 ± 2.66 | 20.91 ± 1.62 | 9.40 | 0 |
| 2859 ± 16 | 27.66 ± 3.55 | 197.8* | 11.40 | 0 |
| 2968 ± 16 | 26.32 ± 1.04 | 15.45 ± 0.33 | 9.95 | 1 |
| 2994 ± 16 | 31.32 ± 2.85 | 5751* | 17.56 | 0 |
| 3011 ± 16 | 10.80 ± 0.70 | 11.76 ± 0.41 | 5.81 | 1 |
| 3292 ± 18 | 51.41 ± 2.40 | 1032* | 22.01 | 0 |
| 3429 ± 19 | 30.15 ± 2.58 | 0.90 ± 0.09 | 0.84 | 1 |
| 3461 ± 19 | 39.16 ± 2.49 | 6.61 ± 0.35 | 4.77 | 1 |
| 3481 ± 19 | 52.31 ± 2.80 | 7197* | 29.30 | 0 |
| 3554 ± 20 | 9.07 ± 3.59 | 10.81 ± 1.12 | 3.46 | 1 |
| 3735 ± 21 | 87.38 ± 3.77 | 91.48* | 26.96 | 0 |
| 3754 ± 21 | 57.91 ± 3.60 | 2571* | 32.17 | 0 |
| 3991 ± 23 | 30.72 ± 1.95 | 61.3 ± 8.1 | 21.31 | 1 |
| 4120 ± 23 | 72.79 ± 4.59 | 54.9 ± 7.2 | 32.13 | 1 |
| 4309 ± 24 | 61.69 ± 4.25 | 21.94 ± 3.41 | 10.26 | 1 |
| 4361 ± 25 | 35.88 ± 4.20 | 2509* | 15.6034 | 0 |
| 4616 ± 27 | 52.06 ± 3.52 | 31.3 ± 3.1 | 13.18 | 1 |
| 4658 ± 27 | 41.23 ± 4.10 | 1172* | 22.89 | 0 |
| 4740 ± 27 | 38.06 ± 3.75 | 43.7 ± 3.5 | 12.05 | 1 |
| 4823 ± 29 | 31.94 ± 3.39 | 76.10* | 11.80 | 0 |
| 5167 ± 30 | 25.00 ± 1.75 | 11.58 ± 4.30 | 8.06 | 1 |
| 5191 ± 30 | 37.77 ± 1.70 | 57.2 ± 5.1 | 15.49 | 1 |
| 5287 ± 30 | 33.00 ± 2.60 | 44.2 ± 3.5 | 10.88 | 1 |
| 5357 ± 31 | 26.10 ± 4.01 | 599* | 11.20 | 0 |
| 5533 ± 33 | 24.71 ± 2.63 | 126.4* | 12.52 | 0 |
| 5838 ± 35 | 67.06 ± 2.78 | 603.1* | 35.50 | 0 |
| 5860 ± 35 | 80.48 ± 4.64 | 88.2 ± 3.6 | 43.41 | 1 |
| 5879 ± 35 | 8.58 ± 1.25 | 68.9 ± 3.5 | 8.05 | 0 |
| 5937 ± 36 | 20.21 ± 1.43 | 327.5 ± 12.2 | 20.15 | 1 |
| 5983 ± 36 | 38.11 ± 2.61 | 286.1* | 19.93 | 0 |
| 6036 ± 36 | 38.89 ± 2.73 | 24.91 ± 1.82 | 8.17 | 1 |
| 6148 ± 37 | 19.98 ± 1.95 | 16.82 ± 1.63 | 6.73 | 1 |
| 6343 ± 38 | 61.84 ± 4.10 | 64.1 ± 4.3 | 14.85 | 1 |
| 6467 ± 38 | 38.90 ± 2.64 | 2089* | 16.88 | 0 |
| 6600 ± 39 | 59.88 ± 4.08 | 753* | 25.31 | 0 |
| 6766 ± 41 | 46.13 ± 3.25 | 50.4 ± 3.5 | 11.21 | 1 |
| 6974 ± 42 | 98.84 ± 6.61 | 601* | 40.34 | 0 |
| 7004 ± 42 | 21.75 ± 2.94 | 441 ± 38 | 9.31 | 0 |
| 7060 ± 43 | 31.13 ± 1.62 | 44.9 ± 3.4 | 12.60 | 1 |
| 7107 ± 43 | 11.47 ± 4.48 | 35.0 ± 3.1 | 4.39 | 1 |
| 7144 ± 44 | 53.29 ± 3.92 | 1300* | 29.30 | 0 |
| 7271 ± 45 | 54.46 ± 4.09 | 24.13 ± 1.98 | 11.99 | 1 |
| 7458 ± 47 | 101.8 ± 5.36 | 1002* | 42.63 | 0 |
| 7561 ± 47 | 33.02 ± 2.60 | 130.3* | 16.26 | 0 |
| 7632 ± 49 | 30.72 ± 1.65 | 24.03 ± 2.53 | 8.62 | 1 |
| 7783 ± 50 | 58.52 ± 5.41 | 35.1 ± 3.7 | 14.80 | 1 |
| 8000 ± 52 | 52.06 ± 14.12 | 1126 ± 143 | 28.51 | 0 |
| 8042 ± 52 | 130.4 ± 9.4 | 1299* | 54.68 | 0 |
| 8142 ± 53 | 17.29 ± 1.44 | 135.2* | 7.16 | 1 |
| 8400 ± 55 | 45.87 ± 3.81 | 228.6* | 18.45 | 0 |
| 8534 ± 56 | 95.33 ± 7.71 | 2491* | 41.02 | 0 |
| 8655 ± 56 | 43.48 ± 3.55 | 372.5* | 18.10 | 0 |
| 8707 ± 57 | 43.49 ± 3.63 | 121.5* | 33.48 | 1 |
| 8910 ± 58 | 21.18 ± 1.79 | 56.3 ± 3.5 | 9.83 | 1 |
| 8970 ± 58 | 56.87 ± 4.71 | 3682* | 24.71 | 0 |

**TABLE VI.** Maxwellian-averaged ($n,\gamma$) cross sections. The values in column 2 correspond to the partial information from the present measurement (1 eV up to 9 keV). The third column contains the MACS averaged over the full thermal spectrum after complementing the measured data by the normalized JENDL-3.3 data between 9 and 500 keV. The MACS at 25 keV (in parentheses) is adopted from Ref. [10] and was used for normalization.

| $kT$ (keV) | P-MACS (mbarn) | MACS (mbarn) |
|---|---|---|
| 5 | 83.5 ± 4.6 | 106.9 ± 5.3 |
| 8 | 45.9 ± 3.0 | 74.8 ± 4.1 |
| 10 | 33.2 ± 2.5 | 63.2 ± 3.9 |
| 12.5 | 23.4 ± 1.9 | 53.9 ± 3.8 |
| 15 | 17.4 ± 1.5 | 48.1 ± 3.8 |
| 17.5 | 13.4 ± 1.1 | 43.8 ± 3.7 |
| 20 | 10.7 ± 0.9 | 40.3 ± 3.3 |
| 25 | 7.2 ± 0.6 | (35.7) |
| 30 | 5.2 ± 0.4 | 32.4 ± 3.1 |
| 35 | 3.9 ± 0.3 | 29.8 ± 3.0 |
| 40 | 3.0 ± 0.3 | 27.7 ± 2.8 |
| 45 | 2.4 ± 0.2 | 25.3 ± 2.5 |
| 50 | 2.0 ± 0.2 | 23.6 ± 2.4 |
| 60 | 1.4 ± 0.1 | 22.0 ± 2.2 |
| 70 | 1.1 ± 0.1 | 19.9 ± 2.0 |
| 85 | 0.7 ± 0.06 | 17.4 ± 1.7 |
| 100 | 0.5 ± 0.04 | 15.2 ± 1.5 |

**TABLE VII. Uncertainties of the lanthanum (*n,γ*) measurement. The statistical error represents the major contribution to the total uncertainty. The statistical uncertainties become larger at higher neutron energy because of lower counting rate, see text for details.**

| Source of uncertainty | Statistical | WF  | Flux | Minor Effects | Total     |
|-----------------------|-------------|-----|------|---------------|-----------|
| Uncertainty (%)       | 4 - 8       | 1.5 | 2    | 0.5           | 4.7 - 8.5 |

**TABLE VIII. Nuclear quantities of the $n + {}^{139}\text{La}$ compound system derived from the observed *s*- and *p*-wave resonances between 0.6 eV and 9 keV.**

| *s*-wave | *p*-wave |
|---|---|
| $\langle \Gamma_\gamma \rangle_{l=0} = 50.7 \pm 5.4$ meV | $\langle \Gamma_\gamma \rangle_{l=1} = 33.6 \pm 6.9$ meV |
| $\langle D \rangle_{l=0} = 252 \pm 22$ eV | $\langle D \rangle_{l=1} < 250$ eV |
| $S_0 = (0.82 \pm 0.05) \times 10^{-4}$ | $S_1 = (0.55 \pm 0.04) \times 10^{-4}$ |
| $\Delta_{3,\text{Exp}} = 0.57 \pm 0.10$ | - |
| $\Delta_{3,\text{DM}} = 0.72 \pm 0.22$ | - |
| $\Delta_{3,\text{GOE}} = 0.60^{+0.16}_{-0.22}$ | - |

**TABLE IX. Selected stars from Figs. 8 and 9 showing differences of more than 0.2 dex in the spectroscopic ratios reported by different authors[a].**

| Star | Ref[b]. | [Fe/H] | [La/Fe] | [Eu/Fe] | [La/Eu] |
|---|---|---|---|---|---|
| HD-2796 | Bu | **− 2.23** | **− 0.22** | − 0.06 | **− 0.16** |
|  | Mc | **− 2.52** | **− 0.48** | 0.12 | **− 0.60** |
| HD-6268 | Mc | − 2.59 | **− 0.03** | 0.68 | **− 0.71** |
|  | Ho | − 2.63 | **0.21** | 0.52 | **− 0.31** |
| HD-29574 | Bu | − 1.81 | **0.38** | **0.76** | − 0.38 |
|  | Jo | − 1.84 | **− 0.18** | **0.20** | − 0.38 |
| HD-74462 | Si | − 1.51 | **0.15** | **0.59** | − 0.45 |
|  | Bu | − 1.56 | **− 0.06** | **0.32** | − 0.38 |
| HD-110184 | Si | − 2.50 | **− 0.16** | 0.26 | − 0.42 |
|  | Bu | − 2.56 | **0.11** | 0.45 | − 0.34 |
| HD-115444 | Jo | **− 2.71** | 0.26 | **0.83** | 0.57 |
|  | Si | **− 3.16** | 0.17 | **0.57** | 0.40 |
| HD-122956 | Si | − 1.69 | 0.02 | **0.37** | − 0.36 |
|  | Bu | − 1.78 | 0.04 | **0.55** | − 0.51 |
| HD-128279 | Si | **− 2.13** | − 0.24 | 0.03 | − 0.28 |
|  | Jo | **− 2.40** | − 0.30 | 0.10 | − 0.40 |
| HD-165169 | Jo | − 2.24 | **− 0.15** | 0.41 | − 0.39 |
|  | Bu | − 2.32 | **0.19** | 0.58 | − 0.49 |
| HD-171496 | Si | − 0.61 | **− 0.04** | **0.22** | − 0.26 |
|  | Bu | − 0.91 | **− 0.47** | **− 0.15** | − 0.32 |
| HD-186478 | Ho | **− 2.61** | 0.07 | 0.61 | − 0.53 |
| | Jo | − **2.42** | 0.01 | 0.54 | − 0.54 |
| HD-204543 | Si | − 1.72 | − **0.01** | **0.21** | − 0.22 |
| | Bu | − 1.84 | **0.27** | **0.46** | − **0.19** |
| HD-206739 | Bu | − 1.58 | − **0.17** | 0.35 | − 0.52 |
| | Si | − 1.61 | **0.10** | 0.46 | − 0.36 |
| HD-232078 | Si | − 1.40 | − **0.17** | 0.15 | − 0.32 |
| | Bu | − 1.54 | **0.12** | 0.40 | − 0.28 |

[a] Discrepant values indicated by bold numbers.

[b] Reference code: Bu [64], Si [65], Mc [66], Jo [67], Ho [68] and Ba [69].

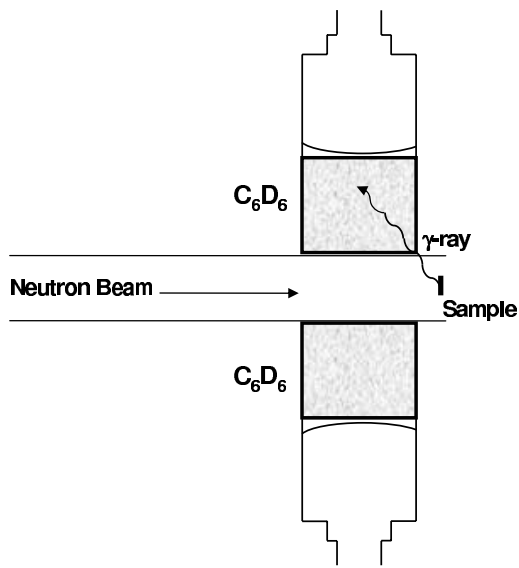

**FIG. 1. Schematic sketch of the n_TOF beam line illustrating the relative positions of sample and detectors.**

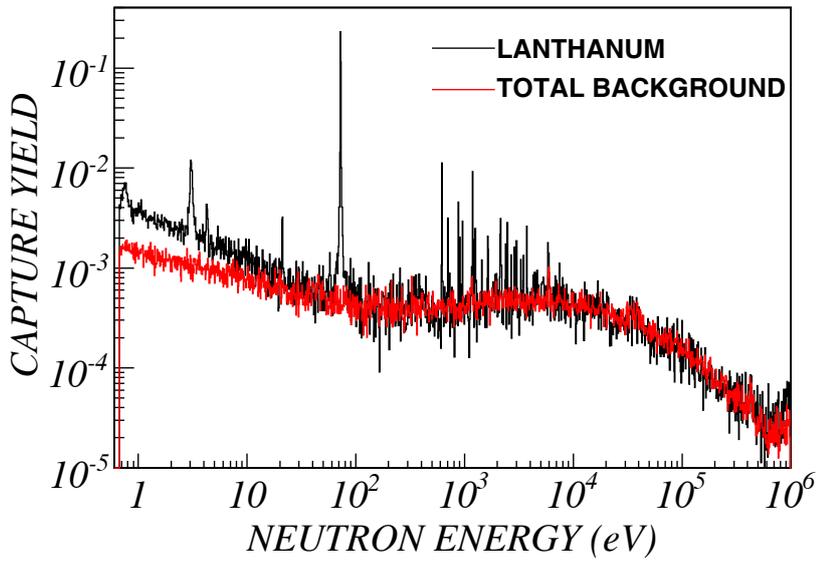

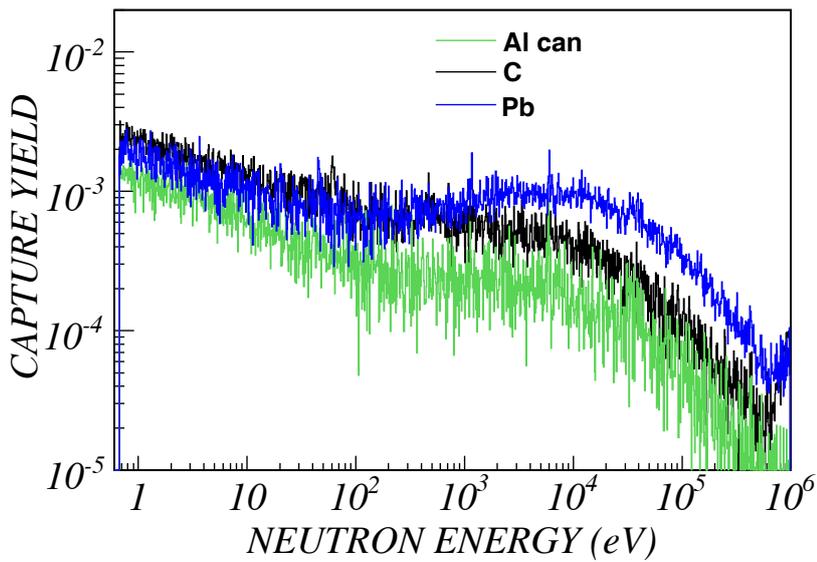

**FIG. 2.** (Color online) Top: Capture yield of the La sample and total background. Bottom: Individual background components: In the keV region the effect of in-beam $\gamma$-rays measured with the Pb sample dominates over the component due to sample scattered neutrons, whereas the contribution of the Al can is always comparably small.

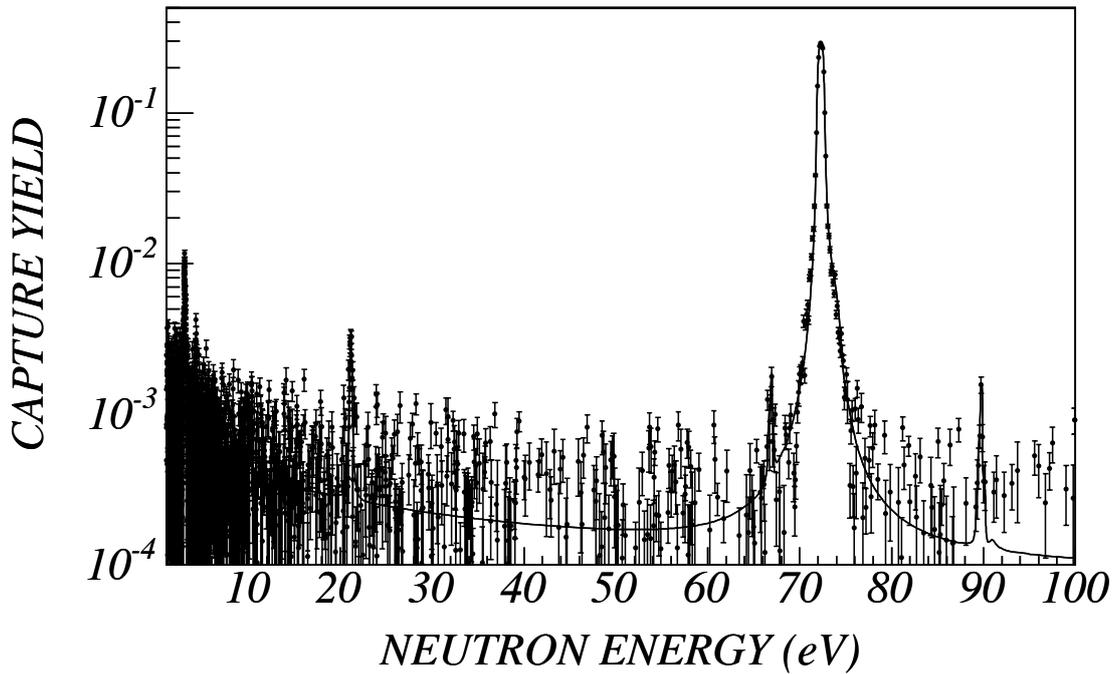

**FIG. 3.** The background subtracted capture yield of the La sample fitted with SAMMY. According to Mughabghab *et al.* [30], the small resonances at 3 eV, 20 eV, 68 eV and 89 eV are due to the minute $^{138}$La impurity of 0.09%.

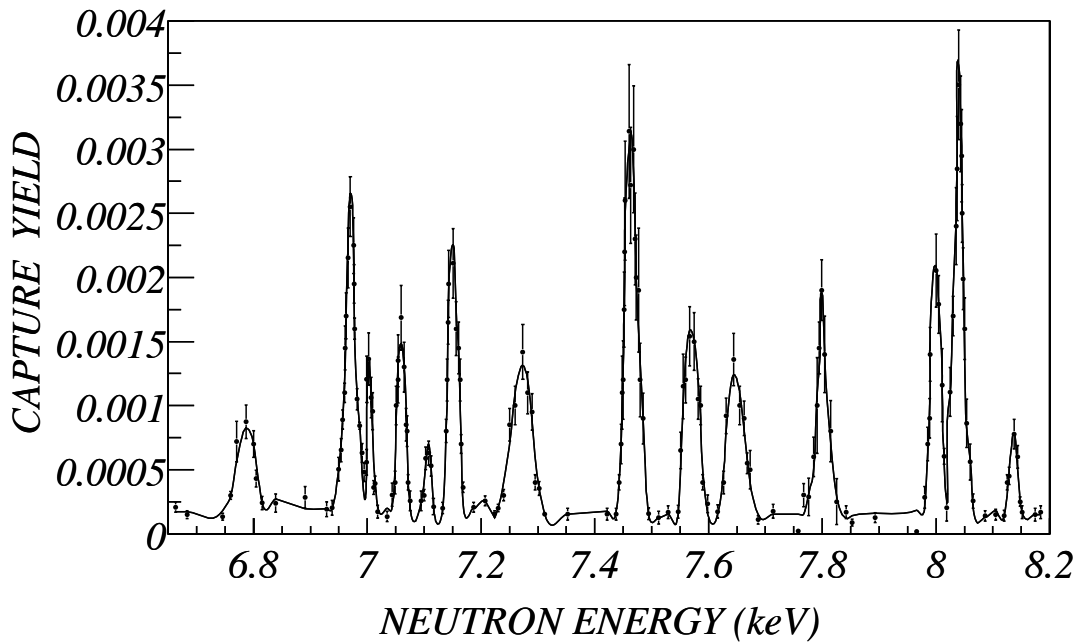

**FIG. 4.** SAMMY fit of the La capture yield in the keV region.

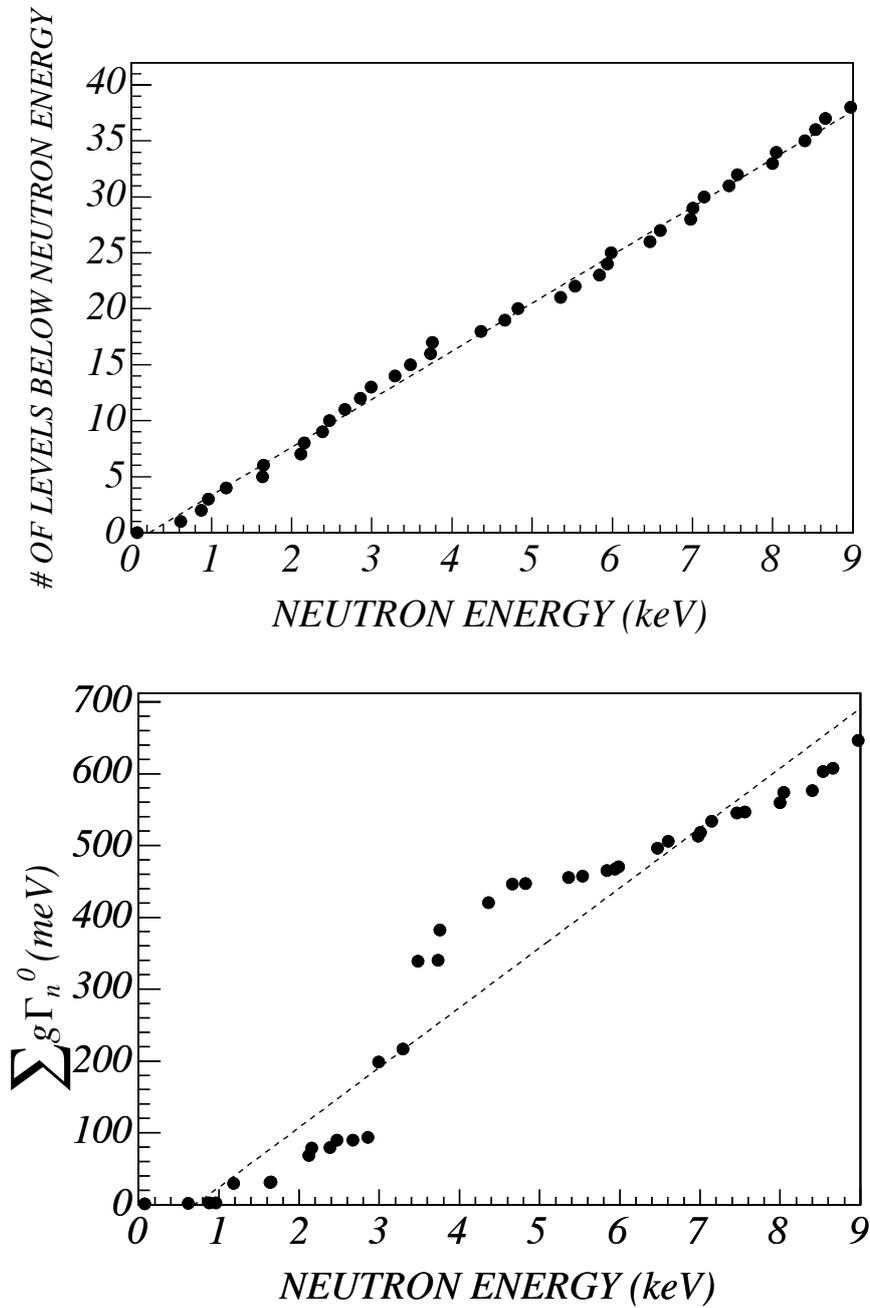

**FIG. 5. Cumulative number of levels (top) and cumulative sum of neutron widths (bottom) of the *s*-wave resonances. The dashed line fitted to the data indicates that the sequence of the *s*-wave levels seems to be complete.**

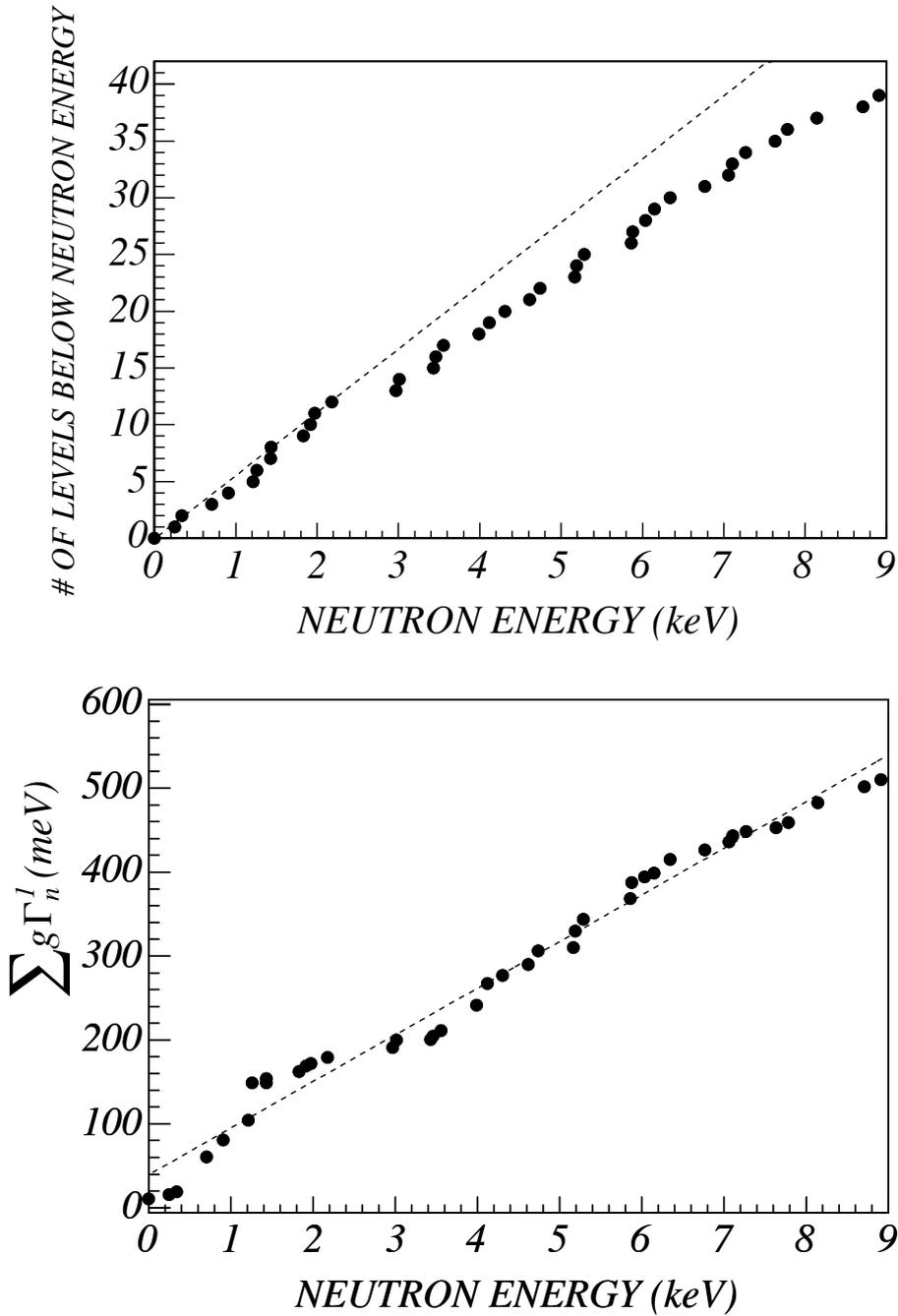

**FIG. 6. Cumulative number of levels (top) and cumulative sum of neutron widths (bottom) of the *p*-wave resonances. The fit to the cumulative number of the levels is limited to the region between 0 and 2.5 keV, where the *p*-wave level sequence seems to be complete. The fit of the cumulative sum of the neutron widths includes the entire energy range.**

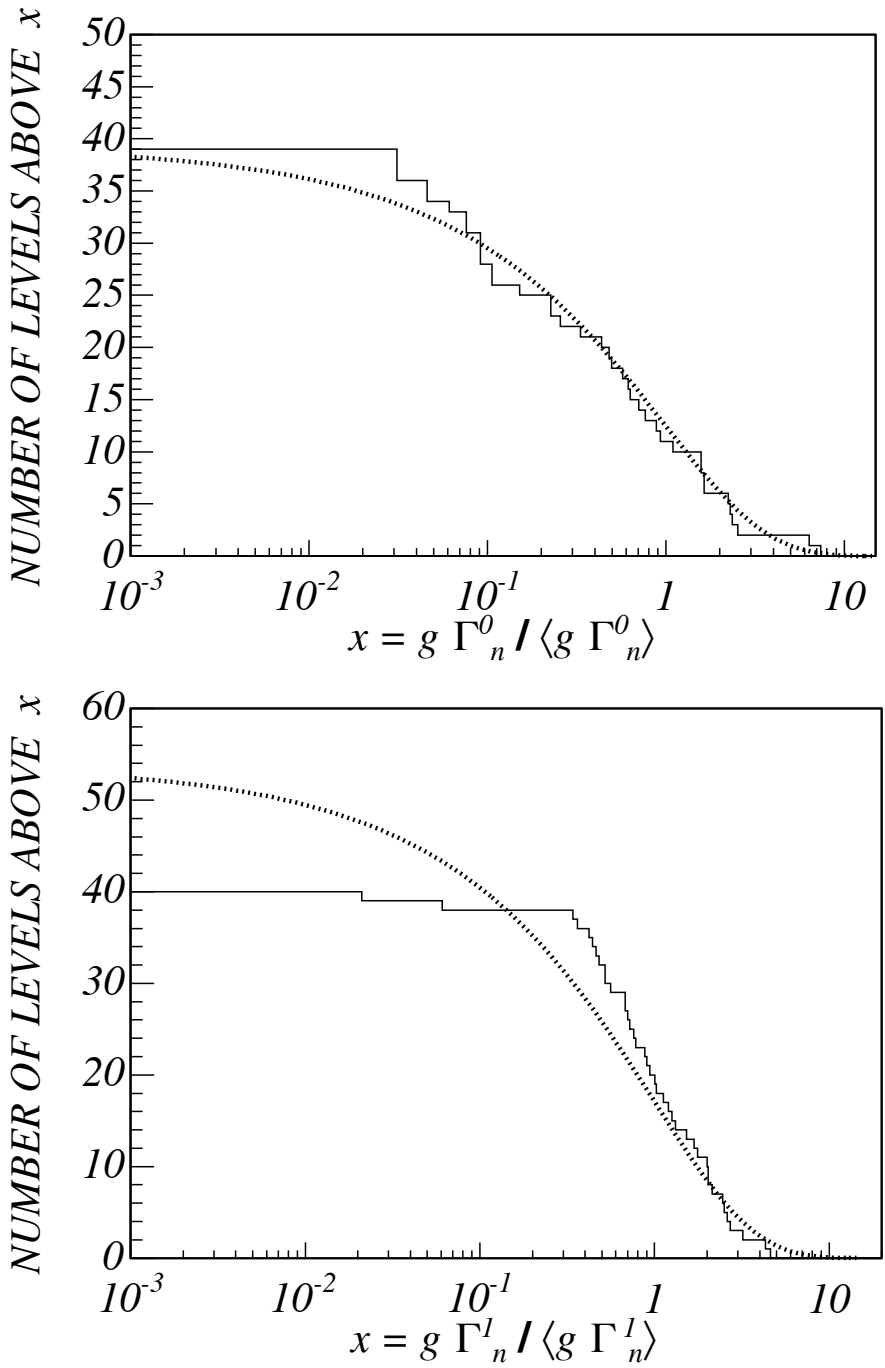

**FIG. 7. Cumulated number of levels with values larger than *x* for *s*-waves (top) and *p*-waves (bottom). The dotted curve indicates the fits via Eq. (7), confirming that the *s*-wave ensemble seems complete while *p*-waves are clearly missing at low values of *x*.**

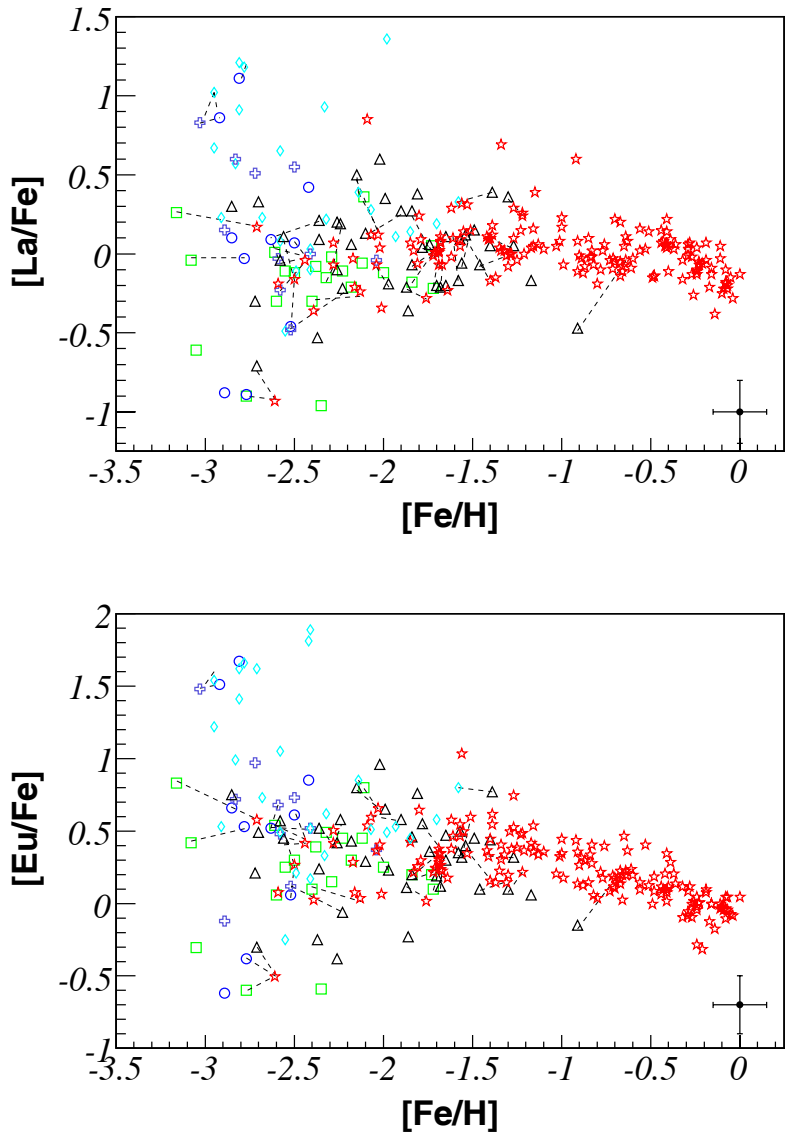

FIG. 8. (Color online) Observed spectroscopic ratios of La (top) and Eu (bottom) with respect to Fe as a function of metallicity. (The references used are denoted by symbols: black triangles [64], red stars [65], blue crosses [66], green squares [67], blue circles [68], and blue diamonds [69]). A typical error bar is shown in the lower right corner. Stars observed by different authors are connected by dashed lines.

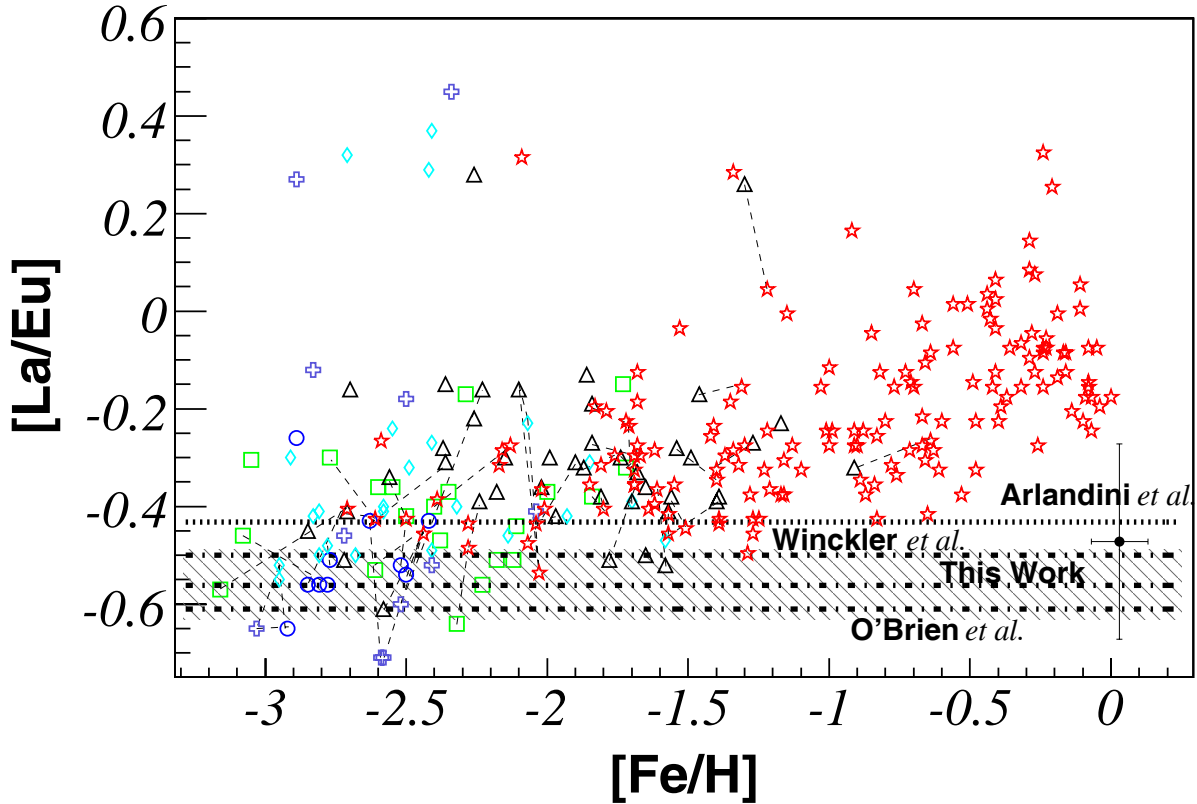

FIG. 9. (Color online) The spectroscopic ratio of La with respect to Eu as a function of metallicity. The symbols are the same as in Fig. 8. The *r*-process ratios calculated according to the TP-AGB model are indicated by horizontal lines, illustrating the result of this work compared to O'Brien *et al.* [10] and Winckler *et al.* [11]. Note, that the older data used by Arlandini *et al.* [50] led to [La/Eu] ratio, which are higher than the observed values at very low metallicity. An indicative and average error bar is shown on the right at bottom. The shaded area represents the total error of the spectroscopic ratio and is induced mainly by the uncertainties of the solar elemental abundances. The stars having high [La/Eu] values (0.2-0.4) represent the *s*-process rich stars and require a different treatment, see text for details.